\documentclass[a4paper,11pt]{article}
\usepackage{a4wide,amsmath,amssymb,bbm}
\usepackage[english]{babel}

\makeatletter

\@addtoreset{equation}{section} \makeatother

\renewcommand{\u}{\underline}

\begin{document}

\hfill{MIFPA-13-13}

\vspace{40pt}

\begin{center}
{\LARGE{\bf The type II superstring to order $\theta^4$}}

\vspace{60pt}

Linus Wulff

\vspace{15pt}

{\it\small George P. \& Cynthia Woods Mitchell Institute for Fundamental Physics and Astronomy,}\\
{\it\small Texas A\&M University, College Station, TX 77843, USA}\\

\vspace{120pt}

{\bf Abstract}

\end{center}

The Green-Schwarz superstring action in a general type IIA or IIB supergravity background is derived up to fourth order in the Grassmann-odd coordinates $\theta$. This is done by solving the superspace Bianchi identities order by order in $\theta$, to quadratic order for all superfields and to quartic order for the supervielbeins. For a large class of backgrounds it is possible to fix the kappa symmetry in such a way that the action actually terminates at the quartic order in $\theta$.

\pagebreak 
%\tableofcontents
\setcounter{page}{1}

%%%%%%%%%%%%%%%%%%%%%%%%%%%%%%%%%%%%%%%%%%%%%%%%%%%%%%%%%%%%%%%%%%%%%%%%

\section{Introduction}
In many examples of the AdS/CFT-correspondence \cite{Maldacena:1997re} string theory on an $AdS$-background with RR-flux is conjectured to be dual to a conformal field theory living on the boundary of $AdS$. Particularly interesting examples, because of their integrable structure, include:
\begin{itemize}
	\item Type IIB string theory on $AdS_5\times S^5$ with RR five-form flux \cite{Bena:2003wd}.
  \item Type IIA string theory on $AdS_4\times\mathbbm{CP}^3$ with RR two- and four-form flux \cite{Arutyunov:2008if,Stefanski:2008ik,Gomis:2008jt}.
  \item Type IIB string theory on $AdS_3\times S^3\times T^4$ or $AdS_3\times S^3\times S^3\times S^1$ with RR (and/or NSNS \cite{Cagnazzo:2012se}) three-form flux \cite{Babichenko:2009dk}.
  \item Type IIA string theory on $AdS_2\times S^2\times T^6$ with RR four-form flux \cite{Sorokin:2011rr}.
\end{itemize}
The last two have both type IIA and type IIB realizations due to T-duality. For the last case there is also a version with both two- and four-form flux also related by T-duality.

An important tool in studying the AdS/CFT-correspondence is the study of semiclassical strings in the relevant background. Since the NSR-formulation of the string is not suited for describing strings in RR backgrounds one typically relies on the Green-Schwarz formulation, which is perfectly suited for describing strings in any supergravity background, at least at the semiclassical level. The backgrounds listed above preserve 32, 24, 16 and 8 supersymmetries respectively and therefore they have a supercoset realization with the same number of fermionic coordinates. For all but the last example this means that the Green-Schwarz superstring, which has 16 physical fermions, can be described by a supercoset sigma model. However, this supercoset sigma model is in general a (partially) kappa symmetry gauge-fixed version of the full Green-Schwarz string which has 32 fermions. Due to subtleties with gauge-fixing of kappa symmetry it turns out that this gauge-fixing is not always compatible with the string configuration one is studying leading to potential problems with the supercoset formulation in certain cases. Because of this it is sometimes desirable to work with the full Green-Schwarz superstring.\footnote{The classical integrability has been extended to the full Green-Schwarz action up to quadratic order in fermions \cite{Sorokin:2010wn,Sundin:2012gc} (and in the $AdS_4$-case to all orders for a truncated model).} Unfortunately the full Green-Schwarz supersting action is only known in a few special backgrounds such as $AdS_5\times S^5$ \cite{Metsaev:1998it}, where it is the same as the supercoset sigma model, $AdS_4\times\mathbbm{CP}^3$ \cite{Gomis:2008jt}, where it was found by dimensional reduction from eleven dimensions, a certain 7-brane background \cite{Russo:1998xv} and plane-wave backgrounds. However, the action is known to quadratic order in fermions in a general type II supergravity background \cite{Cvetic:1999zs}, something which has proved to be quite useful for semiclassical computations.

In this paper we will construct the type II Green-Schwarz superstring action in a general supergravity background (with vanishing gravitino and dilatino) to fourth order in the fermions. We hope that this result will be useful in the study of semiclassical strings in the AdS/CFT-context. To write the Green-Schwarz action to a given order in $\theta$ one must know the supergeometry, specifically the supervielbeins, to the same order. The supergeometry can be found in a systematic way order by order in $\theta$ from the superspace constraints and Bianchi identities. We carry out this construction up to order $\theta^4$ for the supervielbeins.\footnote{An alternative approach would have been to obtain the result by dimensional reduction from eleven dimensions where the supergeometry is known up to order $\theta^5$ \cite{Tsimpis:2004gq}.} It could in principle be pushed to higher orders but the complexity of the expressions increases quite rapidly. In fact, for a large class of backgrounds including the ones mentioned above, the fourth order action is enough in the sense that one can find a certain light-cone kappa symmetry gauge fixing such that the action terminates at the quartic order in fermions. This was suggested in \cite{Sahakian:2004gy} where the kappa symmetry fixed type IIB superstring action was also constructed for these backgrounds. Here we are able to write a more compact and geometrical expression for the action which is not gauge-fixed and covers all possible backgrounds.

The outline of the paper is as follows. In section \ref{sec:action} we give our results for the superstring action in a general type II supergravity background up to quartic order in $\theta$. We also review the argument for why one can find a kappa-gauge such that the action terminates at this order for a large class of interesting backgrounds. The rest of the paper is devoted to the derivation of these results. In section \ref{sec:supergeometry} we outline the systematic process for solving the superspace Bianchi identities order by order in $\theta$. All superfields of the type II background are computed to order $\theta^2$, while the supervielbeins (and $B$ field) are computed to order $\theta^4$. Section \ref{sec:conclusion} contains our conclusions.

Appendix \ref{app:gamma} gives our spinor and gamma matrix conventions and in appendix \ref{sec:IIB} we describe the constraints and Bianchi identities of type IIB supergravity in superspace. For the type IIA case the superspace constraints are listed in appendix \ref{sec:IIA}. The constraints are written in a form which makes it trivial to go between the IIA and the IIB case.

\section{The type II superstring action to order $\theta^4$}\label{sec:action}
The Green-Schwarz superstring action takes the following form in a general type II supergravity background \cite{Grisaru:1985fv}
\begin{equation}
\label{eq:action}
S=-T\int_\Sigma\,\big(\frac12*E^aE^b\eta_{ab}-B\big)\,,
\end{equation}
where $E^a$ ($a=0,\ldots,9$) are the vector supervielbeins of the background pulled back to the worldsheet and $B$ is the NSNS two-form potential also pulled back to the woldsheet. We are using 2d form notation and the Hodge-dual '$*$' is defined using an auxiliary worldsheet metric.\footnote{It is of course also possible to use the Nambu-Goto form of the action, obtained by integrating out the auxiliary worldsheet metric, but then the statement about the action truncating at order $\theta^4$ for certain backgrounds does not apply.} We will assume only that the background is a supergravity solution with vanishing fermionic (gravitino, dilatino) fields. The supervielbeins and $B$ field are superfields and therefore depend both on the ten bosonic coordinates and also on the $32$ fermionic coordinates of the type II superspace. The action therefore has an expansion in even powers of the fermions $\Theta$ up to 32nd order which follows from the expansion of these superfields. In this section we give the expansion up to fourth order in $\Theta$. The rest of the paper is devoted to deriving the expansion of the superfields to this order. We will write the expressions for the IIA case and at the end of the section we explain how to obtain the IIB expressions by simple substitutions in the IIA expressions.

The zeroth order Lagrangian is obtained by simply setting the fermions to zero in (\ref{eq:action}),
\begin{equation}
\mathcal L^{(0)}=\frac12*e^ae^b\eta_{ab}-B^{(0)}\,,
\end{equation}
where we denote the purely bosonic vielbeins by $e^a$ and $B^{(0)}$ is the lowest component in the $\theta$-expansion of $B$. 

The terms quadratic in fermions take the following form
\begin{equation}
\mathcal L^{(2)}=\frac{i}{2}*e^a\,\Theta\Gamma_a\mathcal D\Theta-\frac{i}{2}e^a\,\Theta\Gamma_a\Gamma_{11}\mathcal D\Theta\,,
\end{equation}
where
\begin{equation}
\label{eq:DbA}
\mathcal D\Theta=
\big(d-\frac{1}{4}\omega^{ab}\Gamma_{ab}+\frac{1}{8}e^a\,H_{abc}\,\Gamma^{bc}\Gamma_{11}+\frac{1}{8}e^a\,S\Gamma_a\big)\Theta\,,
\end{equation}
$\omega^{ab}$ is the spin connection, $H=dB$ is the NSNS three-form field strength and
\begin{equation}
\label{eq:SbA}
S=e^\phi\big(\frac12F^{(2)}_{ab}\Gamma^{ab}\Gamma_{11}+\frac{1}{4!}F^{(4)}_{abcd}\Gamma^{abcd}\big)\,.
\end{equation}
The derivative operator $\mathcal D$ is the Killing spinor operator as we will discuss below. The matrix $S$ encodes the dependence on the dilaton $\phi$ and RR-fields (recall that we are describing the IIA case here with RR two- and four-form fields, the IIB case will be discussed below). Here $\Theta$ is a $32$-component Majorana spinor, see Appendix \ref{app:gamma} for our spinor and gamma-matrix conventions. The quadratic action was first derived in \cite{Tseytlin:1996hs,Cvetic:1999zs} starting from the supermembrane action in eleven dimensions.

The main result of this paper is the quartic superstring Lagrangian which takes the form
\begin{eqnarray}
\lefteqn{\mathcal L^{(4)}=}
\nonumber\\
&&{}
-\frac{1}{8}\Theta\Gamma^a*\mathcal D\Theta\,\Theta\Gamma_a\mathcal D\Theta
+\frac{1}{8}\Theta\Gamma^a\mathcal D\Theta\,\Theta\Gamma_a\Gamma_{11}\mathcal D\Theta
+\frac{i}{24}*e^a\,\Theta\Gamma_a\mathcal M\mathcal D\Theta
-\frac{i}{24}e^a\,\Theta\Gamma_a\Gamma_{11}\mathcal M\mathcal D\Theta
\nonumber\\
&&{}
+\frac{i}{3\cdot64}*e^ae^b\,\Theta\Gamma_a(M+\tilde M)S\Gamma_b\Theta
-\frac{i}{3\cdot64}e^ae^b\,\Theta\Gamma_a\Gamma_{11}(M+\tilde M)S\Gamma_b\Theta
\nonumber\\
&&{}
+\frac{1}{3\cdot64}[*e^ce^d\,\Theta\Gamma_c{}^{ab}\Theta-e^ce^d\,\Theta\Gamma_c{}^{ab}\Gamma_{11}\Theta]\,(3\Theta\Gamma_dU_{ab}\Theta
-2\Theta\Gamma_aU_{bd}\Theta)
\nonumber\\
&&{}
-\frac{1}{3\cdot64}[*e^ce^d\,\Theta\Gamma_c{}^{ab}\Gamma_{11}\Theta-e^ce^d\,\Theta\Gamma_c{}^{ab}\Theta]\,(3\Theta\Gamma_d\Gamma_{11}U_{ab}\Theta+2\Theta\Gamma_a\Gamma_{11}U_{bd}\Theta)\,.
\label{eq:L4}
\end{eqnarray}
To shorten the expression we have defined two matrices which are quadratic in fermions
\begin{eqnarray}
\mathcal M^{\u\alpha}{}_{\u\beta}&=&
M^{\u\alpha}{}_{\u\beta}
+\tilde M^{\u\alpha}{}_{\u\beta}
+\frac{i}{8}H_{abc}\,(\Gamma^{ab}\Gamma_{11}\Theta)^{\u\alpha}\,(\Theta\Gamma^c)_{\u\beta}
+\frac{i}{8}H_{abc}\,(\Gamma^{ab}\Theta)^{\u\alpha}\,(\Theta\Gamma^c\Gamma_{11})_{\u\beta}
\nonumber\\
&&{}
+\frac{i}{8}(S\Gamma^a\Theta)^{\u\alpha}\,(\Theta\Gamma_a)_{\u\beta}
-\frac{i}{16}(\Gamma^{ab}\Theta)^{\u\alpha}\,(\Theta\Gamma_aS\Gamma_b)_{\u\beta}
\nonumber\\
M^{\u\alpha}{}_{\u\beta}&=&
\frac12\Theta T\Theta\,\delta^{\u\alpha}_{\u\beta}
-\frac12\Theta\Gamma_{11}T\Theta\,(\Gamma_{11})^{\u\alpha}{}_{\u\beta}
+\Theta^{\u\alpha}\, (T\Theta)_{\u\beta}
+(\Gamma^aT\Theta)^{\u\alpha}\,(\Theta\Gamma_a)_{\u\beta}
\label{eq:M}
\end{eqnarray}
while $\tilde M=\Gamma_{11}M\Gamma_{11}$. In addition two new matrices constructed from the background fields contracted with gamma-matrices appear at this order
\begin{eqnarray}
T&=&\frac{i}{2}\nabla_a\phi\,\Gamma^a+\frac{i}{24}H_{abc}\,\Gamma^{abc}\Gamma_{11}+\frac{i}{16}\Gamma_aS\Gamma^a
\label{eq:T}
\\
\nonumber\\
U_{ab}&=&
\frac{1}{4}\nabla_{[a}H_{b]cd}\,\Gamma^{cd}\Gamma_{11}
+\frac{1}{4}\nabla_{[a}S\Gamma_{b]}
-\frac{1}{4}(R_{abcd}+\frac12H_{ace}H_{bd}{}^e)\,\Gamma^{cd}
\nonumber\\
&&{}
+\frac{1}{32}S\Gamma_{[a}S\Gamma_{b]}
-\frac{1}{32}H_{cd[a}\,(S\Gamma_{b]}\Gamma^{cd}+\Gamma^{cd}S\Gamma_{b]})\Gamma_{11}\,.
\label{eq:Uab}
\end{eqnarray}
These have a simple interpretation as the matrices that appear in the conditions that ensure supersymmetry of the background. For the background to preserve some supersymmetry the corresponding supersymmetry parameters should make the variation of the dilatino and the gravitino field strength vanish. These conditions read
\begin{eqnarray}
0&=&\delta\chi_{\u\alpha}=\epsilon^{\u\beta}\nabla_{\u\beta}\chi_{\u\alpha}|_{\Theta=0}=(T\epsilon)_{\u\alpha}
\nonumber\\
0&=&\delta\psi_{ab}^{\u\alpha}=\epsilon^{\u\beta}\nabla_{\u\beta}\psi_{ab}^{\u\alpha}|_{\Theta=0}=(U_{ab}\epsilon)^{\u\alpha}
\label{eq:SUSYcond}
\end{eqnarray}
where we used (\ref{eq:dalphachiA}) and (\ref{eq:dalphapsiA}). Therefore supersymmetries of the background correspond to spinors annihilated by the matrices $T$ and $U_{ab}$, i.e. $T$ and $U_{ab}$ are typically proportional to projection operators $(1-\mathcal P)$ where $\mathcal P$ projects on the supersymmetries of the background. Normally one talks about the supersymmetry variation of the gravitino itself vanishing which is equivalent to the Killing spinor equation
\begin{equation}
\mathcal D\epsilon=0\,.
\end{equation}
The equation for the vanishing of the supersymmetry variation of the gravitino field strength is in fact the integrability condition for the Killing spinor equation since
\begin{equation}
\mathcal D\epsilon=0\qquad\Rightarrow\qquad0=\mathcal D^2\epsilon=\frac12e^be^aU_{ab}\epsilon\,.
\end{equation}
We therefore see that the matrices $T$ and $U_{ab}$ have nice interpretations in the type II supergravity. When the background preserves some supersymmetry or, more generally, has some superisometries the string action will be invariant under the superisometry transformations 
\begin{equation}
\delta\Theta^\alpha=\Xi^\alpha(x,\Theta)\,,\qquad\delta x^m=K^m(x,\Theta)\,.
\end{equation}
The superfields $K^m$ and $\Xi^\alpha$ can be constructed order by order in $\theta$ in a very similar way to how the supervielbeins are constructed (see section \ref{sec:supergeometry}). For a supersymmetry the lowest component of $\Xi$ is the Killing spinor while the lowest component of $K$ vanishes.

So far all that we have said applies to a type IIA supergravity background, however we have chosen to write the expressions in such a way that they generalize almost trivially to the type IIB case. The $32$-component Majorana spinor $\Theta^{\u\alpha}$ should be replace by a doublet of $16$-component Majorana-Weyl spinors $\Theta^{\alpha i}$ $i=1,2$. Similarly the gamma-matrices are replaced as follows
\begin{equation}
\Gamma_a\rightarrow\gamma_a\,,\qquad\Gamma_{11}\rightarrow\sigma^3\qquad(\mbox{except:}\quad\Gamma_{11}T\rightarrow-\sigma^3T)\,.
\end{equation}
Finally, instead of the $S$ defined in (\ref{eq:SbA}) one should use the expression appropriate to type IIB
\begin{equation}
\label{eq:SbB}
S=-e^\phi\big(\varepsilon\gamma^a F^{(1)}_a+\frac{1}{3!}\sigma^1\gamma^{abc}F^{(3)}_{abc}+\frac{1}{2\cdot5!}\varepsilon\gamma^{abcde} F^{(5)}_{abcde}\big)\,.
\end{equation}
Here $(\sigma^1,\sigma^2=-i\varepsilon,\sigma^3)$ are Pauli matrices and $\gamma_a$ are $16$-component gamma-matrices defined in Appendix \ref{app:gamma}. With these replacements all the previous expressions apply also for the superstring in a type IIB supergravity background.

\subsection{Exact quartic action for certain backgrounds}
As was pointed out in \cite{Sahakian:2004gy} many supergravity backgrounds of interest belong to a class for which the fourth order action is actually the complete answer. This class includes for example the backgrounds listed in the beginning of the introduction. This statement is true provided that one fixes the kappa-symmetry of the string in the appropriate way. The argument is quite simple: Suppose it is possible to find light-cone coordinates such that
\begin{itemize}
	\item[1.] The supergravity fields depend only on the transverse coordinates, \emph{i.e.} $\nabla_\pm\phi=0$, $\nabla_\pm H_{abc}=0$ etc.
  \item[2.] The background tensor fields have only transverse indices or a pair of $+-$ indices, \emph{i.e.} $R_{\pm a'b'c'}=0$, $F_{\pm a'_1\cdots a'_n}=0$ etc. (the prime denotes transverse directions).
\end{itemize}
Then, fixing the kappa-symmetry by demanding that $\Gamma^+\Theta=0$, the only non-zero spinor bilinears that can appear in the string action are of the form
\begin{equation}
\Theta\Gamma_{a'_1\cdots a'_n}\Gamma^-\mathcal D\Theta\qquad\mbox{or}\qquad\Theta\Gamma_{a'_1\cdots a'_n}\Gamma^-\Theta\,.
\end{equation}
Because of the assumptions (1) and (2) above the single '$-$' index can only be absorbed by multiplying with a vielbein $e^+$. This means that in the Lagrangian (\ref{eq:action}) each spinor bilinear must be accompanied by a vielbein and since the Lagrangian contains at most two vielbeins the action must truncate at fourth order in $\Theta$.

For these backgrounds the action presented here is therefore the complete answer (at least in this kappa-gauge). This makes the quartic action particularly interesting. Note that the statements here apply to the action after fixing kappa symmetry but \emph{before} fixing the bosonic symmetries. Fixing also the bosonic symmetries will typically reintroduce higher order $\theta$-terms in the action through the Virasoro constraints. It is known that this can be avoided in certain cases including the first three backgrounds listed in the introduction by choosing a special, non-conformal, ''$AdS$ light-cone'' gauge \cite{Metsaev:2000yf,Metsaev:2000yu}.\footnote{I want to thank A. Tseytlin for useful discussions of the gauge-fixing.} Note also that the gauge $\Gamma^+\Theta=0$ is not automatically compatible with a certain bosonic gauge-fixing for some string configuration. The consistency of the gauge-fixing must be checked by hand.

\section{Solving the Bianchi identities for the $\theta$ dependence}\label{sec:supergeometry}
In this section we show how to obtain the type II supergeometry systematically as an expansion in $\theta$. All superfields are obtained to order $\theta^2$ while the supervielbeins and $B$-field, which are needed for the string action, are obtained to order $\theta^4$. It is in principle straightforward to push the calculation to higher orders in $\theta$ but the expressions become quite long and not very illuminating. We assume that the fermionic fields are zero in the background. It is straightforward to include them but the expressions become longer and we did not find it particularly useful.

The method we use is essentially the superspace normal coordinate expansion outlined in \cite{Atick:1986jr}, applied to the fermionic coordinates $\theta$ (see \cite{Tsimpis:2004gq} for an application of this procedure to eleven-dimensional supergravity). Rescaling the fermions by a parameter $t$
\begin{equation}
\theta\rightarrow t\theta
\end{equation} 
the $\theta$-expansion of a superfield becomes an expansion in $t$, \emph{e.g.}
\begin{equation}
E^a=E^{(0)\,a}+t^2E^{(2)\,a}+t^4E^{(4)\,a}+\ldots
\end{equation}
for the vector supervielbein. One can then write a set of first order (coupled) differential equations in the variable $t$ for the set of superfields of type II supergravity. To do this one uses the fact that
\begin{equation}
\frac{d}{dt}=\Theta^{\u\alpha}\nabla_{\u\alpha}\,,
\end{equation}
where the normal coordinate $\Theta$ is defined as
\begin{equation}
\label{eq:normalTheta}
\Theta^{\u\alpha}=\theta^{\u\mu}E_{\u\mu}{}^{\u\alpha}=i_\theta E^{\u\alpha}\,,\qquad i_\theta E^a=0\,,\qquad i_\theta\Omega^{ab}=0\,.
\end{equation}
For the supervielbeins one then finds for example
\begin{equation}
\frac{d}{dt}E^A=i_\theta dE^A+di_\theta E^A=\mathcal L_\theta E^A\qquad A=(\u\alpha,a)
\end{equation}
and similarly for the spin connection superfield $\Omega^{ab}$. Using the definition of torsion and curvature (see the Appendix), together with (\ref{eq:normalTheta}) one finds
\begin{eqnarray}
\frac{d}{dt}E^a&=&i_\theta dE^a=i_\theta T^a\\
\frac{d}{dt}E^{\u\alpha}&=&d\Theta^{\u\alpha}+i_\theta dE^{\u\alpha}=([d-\frac{1}{4}\Omega^{ab}\Gamma_{ab}]\Theta)^{\u\alpha}+i_\theta T^{\u\alpha}\\
\frac{d}{dt}\Omega^{ab}&=&i_\theta R^{ab}\,,
\end{eqnarray}
where $T^a$, $T^{\u\alpha}$ are the components of the torsion superfield and $R_{ab}$ is the Riemann curvature superfield. These fields are subject to superspace constraints which are described in the Appendix (as in the previous section we write the equations appropriate to type IIA and describe how to obtain the type IIB case by simple substitutions at the end of the section), \emph{e.g}
\begin{equation}
T^a=-\frac{i}{2}E\Gamma^aE\,.
\end{equation}
Using these constraints, specifically (\ref{eq:TalphaAhalf}), (\ref{eq:TalphaAone}), (\ref{eq:RAone}) and (\ref{eq:dim-threehalfRA}) the equations become
\begin{eqnarray}
\frac{d}{dt}E^a&=&-iE\Gamma^a\Theta\label{eq:ddtEa}\\
\frac{d}{dt}E^{\u\alpha}&=&([d-\frac{1}{4}\Omega^{ab}\Gamma_{ab}]\Theta)^{\u\alpha}
+\frac{1}{8}E^a\,H_{abc}\,(\Gamma^{bc}\Gamma_{11}\Theta)^{\u\alpha}
+\frac{1}{8}E^a\,(S\Gamma_a\Theta)^{\u\alpha}
\nonumber\\
&&{}
-\frac12\Theta^{\u\alpha}\,E\chi
+\frac12(\Gamma_{11}\Theta)^{\u\alpha}\,E\Gamma_{11}\chi
+\frac12E^{\u\alpha}\,\Theta\chi
-\frac12(\Gamma_{11}E)^{\u\alpha}\,\Theta\Gamma_{11}\chi
\nonumber\\
&&{}
-\frac12(\Gamma_a\chi)^{\u\alpha}\,E\Gamma^a\Theta
+\frac12(\Gamma_a\Gamma_{11}\chi)^{\u\alpha}\,E\Gamma^a\Gamma_{11}\Theta
\label{eq:ddtEalpha}\\
\frac{d}{dt}\Omega^{ab}&=&
-\frac{i}{2}H^{abc}\,\Theta\Gamma_c\Gamma_{11}E
+\frac{i}{4}\Theta\Gamma^{[a}S\Gamma^{b]}E
+\frac{i}{2}E^c\,\Theta\Gamma_c\psi^{ab}
-iE^c\,\Theta\Gamma^{[a}\psi^{b]}{}_c\,.
\label{eq:ddtOmega}
\end{eqnarray}
The bosonic superfields appearing in these expressions are the NSNS three-form field $H_{abc}$ and the matrix encoding the RR-superfields $S$, defined in (\ref{eq:SA}) and (\ref{eq:SB}) for the type IIA and IIB case respectively, and the fermionic superfields are the dilatino $\chi$ and gravitino field strength $\psi_{ab}$.

Besides the equations for the $\theta$-expansion of the supervielbeins (\ref{eq:ddtEa}), (\ref{eq:ddtEalpha}) and the spin connection (\ref{eq:ddtOmega}) we also need the corresponding equations for the remaining superfields of the theory which appear in the right-hand-sides of these equations. These are easily obtained, for example, from (\ref{eq:dalphaHA}) we find
\begin{equation}
\label{eq:ddtH}
\frac{d}{dt}H_{abc}=\Theta^{\u\alpha}\nabla_{\u\alpha}H_{abc}=3i\Theta\Gamma_{[a}\Gamma_{11}\psi_{bc]}\,.
\end{equation}
From the definition of the dilatino superfield, $\chi_{\u\alpha}=\nabla_{\u\alpha}\phi$, we get
\begin{equation}
\label{eq:ddtphi}
\frac{d}{dt}\phi=\Theta\chi\,.
\end{equation}
Similarly, using the equation for the spinor derivative of $S$ (\ref{eq:dalphaSA}), we find
\begin{eqnarray}
\lefteqn{\frac{d}{dt}\tilde S^{\u{\beta\gamma}}
=
S^{\u{\beta\gamma}}\,\chi\Theta
+\Theta^{[\u\beta}\,(S\chi)^{\u\gamma]}
-(\Gamma_{11}\Theta)^{[\u\beta}\,(\Gamma_{11}S\chi)^{\u\gamma]}
+(S\Gamma^b\Theta)^{[\u\beta}\,(\Gamma_b\chi)^{\u\gamma]}
}
\nonumber\\
&&{}
+(S\Gamma^b\Gamma_{11}\Theta)^{[\u\beta}\,(\Gamma_{11}\Gamma_b\chi)^{\u\gamma]}
-2i(\Gamma^{cd}\Theta)^{[\u\beta}\,(\psi_{cd})^{\u\gamma]}
+2i(\Gamma^{cd}\Gamma_{11}\Theta)^{[\u\beta}\,(\Gamma_{11}\psi_{cd})^{\u\gamma]}\,.
\label{eq:ddtS}
\end{eqnarray}
where $\tilde S=\Gamma_{11}S\Gamma_{11}$ which is equal to $S$ in the IIA case while in the IIB case we have $\tilde S=\sigma^3S\sigma^3=-S$, hence the notation. The corresponding equation for the Riemann tensor superfield, although it will not be needed to find the superstring action to order $\theta^4$, is
\begin{eqnarray}
\frac{d}{dt}R_{ab}{}^{cd}&=&
-i\Theta\Gamma_{[a}\nabla_{b]}\psi^{cd}
+\frac{i}{8}H_{ef[a}\,\Theta\Gamma^{ef}\Gamma_{b]}\Gamma_{11}\psi_{cd}
+\frac{i}{4}H_{[a}{}^{e[c}\,\Theta\Gamma_e\Gamma_{11}\psi_{b]}{}^{d]}
\nonumber\\
&&{}
+\frac{i}{8}\Theta\Gamma_{[a}S\Gamma_{b]}\psi^{cd}
+\frac{i}{4}\Theta\Gamma_{[a}S\Gamma^{[c}\psi_{b]}{}^{d]}
+[(ab)\leftrightarrow(cd)]\,.
\label{eq:ddtR}
%
%\nabla_{[\a}R_{ab]cd}+T_{[\a a}{}^BR_{|B|b]cd}=0
%\nabla_\a R_{abcd}=2\nabla_{[a}R_{\a b]cd}-2T_{\a[a}{}^\b R_{|\b|b]cd}-R_{\a\b cd}T_{ab}{}^\b
%\nabla_{[a}\psi_{bc]}^\b=T_{[c|\c|}{}^\b \psi_{ab]}{}^\c
\end{eqnarray}

Finally we need the equations for the fermionic superfields. Using (\ref{eq:dalphachiA}) we find for the dilatino
\begin{equation}
\label{eq:ddtchi}
\frac{d}{dt}\chi_{\u\alpha}
=
\frac{i}{2}\nabla_a\phi\,(\Gamma^a\Theta)_{\u\alpha}
+\frac{i}{24}H_{abc}\,(\Gamma^{abc}\Gamma_{11}\Theta)_{\u\alpha}
+\frac{i}{16}(\Gamma^aS\Gamma_a\Theta)_{\u\alpha}
+\frac12\chi_{\u\alpha}\,\Theta\chi
+\frac12(\Gamma_{11}\chi)_{\u\alpha}\,\Theta\Gamma_{11}\chi\,,
\end{equation}
while for the gravitino field strength we find, using (\ref{eq:dalphapsiA}),
\begin{eqnarray}
\frac{d}{dt}\psi_{ab}^{\u\alpha}&=&
\frac{1}{4}\nabla_{[a}H_{b]cd}\,(\Gamma^{cd}\Gamma_{11}\Theta)^{\u\alpha}
+\frac{1}{4}(\nabla_{[a}S\Gamma_{b]}\Theta)^{\u\alpha}
-\frac{1}{4}(R_{abcd}+\frac12H_{ace}H_{bd}{}^e)(\Gamma^{cd}\Theta)^{\u\alpha}
\nonumber\\
&&{}
-\frac{1}{32}H_{cd[a}\,(S\Gamma_{b]}\Gamma^{cd}\Gamma_{11}\Theta)^{\u\alpha}
-\frac{1}{32}H_{cd[a}\,(\Gamma^{cd}S\Gamma_{b]}\Gamma_{11}\Theta)^{\u\alpha}
+\frac{1}{32}(S\Gamma_{[a}S\Gamma_{b]}\Theta)^{\u\alpha}
\nonumber\\
&&{}
-\frac12\Theta^{\u\alpha}\,\psi_{ab}\chi
+\frac12(\Gamma_{11}\Theta)^{\u\alpha}\,\psi_{ab}\Gamma_{11}\chi
+\frac12\psi_{ab}^{\u\alpha}\,\Theta\chi
-\frac12(\Gamma_{11}\psi_{ab})^{\u\alpha}\,\Theta\Gamma_{11}\chi
\nonumber\\
&&{}
+\frac12(\Gamma_c\chi)^{\u\alpha}\,\Theta\Gamma^c\psi_{ab}
-\frac12(\Gamma_c\Gamma_{11}\chi)^{\u\alpha}\,\Theta\Gamma^c\Gamma_{11}\psi_{ab}\,.
\label{eq:ddtpsiab}
\end{eqnarray}
This completes the list of superfields in the theory, but since we are interested in the string action it is not enough to know the NSNS three-form field strength $H$, we want to know its two-form potential $B$ which appears in the string action. Fortunately this is easily obtained from $H$ by the formula
\begin{equation}
B=B^{(0)}+\int_0^1dt\,i_\theta H\,,
\end{equation}
where $B^{(0)}$ is the purely bosonic part of $B$. Using the superspace constraint on $H$ in (\ref{eq:dimzeroH}) we find
\begin{equation}
\label{eq:Beqn}
B=B^{(0)}-i\int_0^1dt\,E^c\,E\Gamma_c\Gamma_{11}\Theta\,,
\end{equation}
which is easily evaluated once we know the form of the supervielbeins to a given order. Analogous formulas can be used to obtain the RR potentials which are needed if one wants to write down D-brane actions.

In general this set of coupled first order differential equations has to be solved order by order in $t$, i.e. $\theta$. In very special cases, namely when the background is maximally supersymmetric they can however be solved exactly in a rather simple closed form. The reason for this is that for a maximally supersymmetric background the fermionic superfields $\chi$ and $\psi_{ab}$ vanish identically (at lowest order this follows from the supersymmetry transformations (\ref{eq:SUSYcond})), leading to drastic simplifications of the system of equations. This simplification is related to the fact that in the maximally supersymmetric case the superspace is in fact a supercoset space. This simplifying structure has been exploited to construct the full supergeometry of $AdS_5\times S^5$ \cite{Metsaev:1998it,Kallosh:1998zx} and also those of $AdS_4\times S^7$ and $AdS_7\times S^4$ \cite{Kallosh:1998zx,deWit:1998yu} in eleven dimensions. When there is less than maximal supersymmetry a subspace of the superspace can still be a supercoset. This leads to some simplification but it is not clear if this is enough to find the full supergeometry directly instead of order by order in $\theta$.

As in the previous section we have written the equations in such a way that the corresponding type IIB equations can be obtained by trivial substitutions. These are as follows
\begin{equation}
\Gamma_a\rightarrow\gamma_a\,,\qquad\Gamma_{11}\rightarrow\sigma^3\qquad(\mbox{except:}\quad\Gamma_{11}\chi\rightarrow-\sigma^3\chi)
\end{equation}
and using the appropriate form of $S$ given in (\ref{eq:SB}).

\subsection{Solution to order $\theta^2$}
Here we construct all the superfields of the supergravity background up to order $\theta^2$ using the approach laid out in the previous section. We will assume that in the supergravity background the fermionic fields vanish. This means that the fermionic superfields will have an expansion in odd powers of $\theta$ and the bosonic ones will have an expansion in even powers of $\theta$.

For the supervielbeins we find, using (\ref{eq:ddtEalpha}) and (\ref{eq:ddtEa}),
\begin{eqnarray}
E^{(1)\,\u\alpha}=(\mathcal D\Theta)^{\u\alpha}\,,\qquad E^{(2)\,a}=\frac{i}{2}\Theta\Gamma^a\mathcal D\Theta\,,
\end{eqnarray}
where $\mathcal D$ is the Killing spinor derivative operator defined in (\ref{eq:DbA}). The dilatino and gravitino field strength superfields at linear order in $\theta$ are easily found by evaluating (\ref{eq:ddtchi}) and (\ref{eq:ddtpsiab}) at $\theta=0$ and one obtains
\begin{equation}
\chi^{(1)}_{\u\alpha}=(T\Theta)_{\u\alpha}\,,\qquad\psi_{ab}^{(1)\,\u\alpha}=(U_{ab}\Theta)^{\u\alpha}
\end{equation}
where the matrices $T$ and $U_{ab}$ are defined in (\ref{eq:T}) and (\ref{eq:Uab}) and determine the amount of supersymmetry of the background. Once we know all fermionic superfields at linear order it is easy to find the bosonic ones at the quadratic order. From (\ref{eq:ddtphi}) and (\ref{eq:ddtH}) we find for the dilaton and NSNS three-form
\begin{equation}
\phi^{(2)}=\frac12\Theta T\Theta\,,\qquad H^{(2)}_{abc}=\frac{3i}{2}\Theta\Gamma_{[a}\Gamma_{11}U_{bc]}\Theta\,,
\end{equation}
while from (\ref{eq:ddtS}) we get
\begin{eqnarray}
\lefteqn{\tilde S^{(2)\,\u{\beta\gamma}}
=
\frac12\tilde S^{\u{\beta\gamma}}\,\Theta T\Theta
+\frac12\Theta^{[\u\beta}\,(ST\Theta)^{\u\gamma]}
-\frac12(\Gamma_{11}\Theta)^{[\u\beta}\,(\Gamma_{11}ST\Theta)^{\u\gamma]}
+\frac12(S\Gamma^a\Theta)^{[\u\beta}\,(\Gamma_aT\Theta)^{\u\gamma]}
}
\nonumber\\
&&{}
-\frac12(\Gamma_{11}S\Gamma^a\Theta)^{[\u\beta}\,(\Gamma_{11}\Gamma_aT\Theta)^{\u\gamma]}
-i(\Gamma^{ab}\Theta)^{[\u\beta}\,(U_{ab}\Theta)^{\u\gamma]}
+i(\Gamma^{cd}\Gamma_{11}\Theta)^{[\u\beta}\,(\Gamma_{11}U_{cd}\Theta)^{\u\gamma]}\,.
\end{eqnarray}
Finally, from (\ref{eq:ddtOmega}) and (\ref{eq:ddtR}) we find the spin connection and Riemann curvature tensor
\begin{eqnarray}
\Omega^{(2)\,ab}=
-\frac{i}{4}H^{abc}\,\Theta\Gamma_c\Gamma_{11}\mathcal D\Theta
+\frac{i}{8}\Theta\Gamma^{[a}S\Gamma^{b]}\mathcal D\Theta
+\frac{i}{4}e^c\,\Theta\Gamma_cU^{ab}\Theta
-\frac{i}{2}e^c\,\Theta\Gamma^{[a}U^{b]}{}_c\Theta
\end{eqnarray}
and
\begin{eqnarray}
R^{(2)}_{ab}{}^{cd}&=&
-\frac{i}{2}\Theta\Gamma_{[a}\nabla_{b]}U^{cd}\Theta
+\frac{i}{16}H_{ef[a}\,\Theta\Gamma^{ef}\Gamma_{b]}\Gamma_{11}U_{cd}\Theta
+\frac{i}{8}H_{[a}{}^{e[c}\,\Theta\Gamma_e\Gamma_{11}U_{b]}{}^{d]}\Theta
\nonumber\\
&&{}
+\frac{i}{16}\Theta\Gamma_{[a}S\Gamma_{b]}U^{cd}\Theta
+\frac{i}{8}\Theta\Gamma_{[a}S\Gamma^{[c}U_{b]}{}^{d]}\Theta
+[(ab)\leftrightarrow(cd)]\,.
\end{eqnarray}
Using (\ref{eq:Beqn}) the $B$-field follows from the expression for the supervielbeins and we get
\begin{equation}
B^{(2)}=\frac{i}{2}e^c\,\Theta\Gamma_c\Gamma_{11}\mathcal D\Theta\,.
\end{equation}
This completes the construction of all superfields to order $\theta^2$. The corresponding expressions for the type IIB case are obtained by simple substitutions as explained at the end of section \ref{sec:action}.

\subsection{Supervielbeins and $B$-field to order $\theta^4$}
In this section we will derive the supervielbeins and $B$-field, which are needed to write the superstring action, to order $\theta^4$. The other superfields can also be easily obtained but we will not give them since we do not need them. From (\ref{eq:ddtEa}) it follows that the fourth order terms in the vector supervielbein are given by
\begin{equation}
E^{(4)\,a}=\frac{i}{4}\Theta\Gamma^aE^{(3)}\,.
\end{equation}
And from (\ref{eq:Beqn}) it follows that the $B$-field at this order is given by
\begin{equation}
B^{(4)}=-\frac{1}{8}\Theta\Gamma^a\mathcal D\Theta\,\Theta\Gamma_a\Gamma_{11}\mathcal D\Theta
+\frac{i}{4}e^c\,\Theta\Gamma_c\Gamma_{11}E^{(3)}\,.
\end{equation}
All that remains then is to determine the cubic terms in the spinor supervielbein. From (\ref{eq:ddtEalpha}) on finds
\begin{eqnarray}
\lefteqn{E^{(3)\,\u\alpha}=
-\frac{1}{12}\Omega^{(2)ab}(\Gamma_{ab}\Theta)^{\u\alpha}
+\frac{1}{24}E^{(2)\,a}\,H_{abc}\,(\Gamma^{bc}\Gamma_{11}\Theta)^{\u\alpha}
+\frac{1}{24}e^a\,H^{(2)}_{abc}\,(\Gamma^{bc}\Gamma_{11}\Theta)^{\u\alpha}
}
\nonumber\\
&&{}
+\frac{1}{24}E^{(2)\,a}\,(S\Gamma_a\Theta)^{\u\alpha}
+\frac{1}{24}e^a\,(S^{(2)}\Gamma_a\Theta)^{\u\alpha}
-\frac16\Theta^{\u\alpha}\,E^{(1)}\chi^{(1)}
+\frac16(\Gamma_{11}\Theta)^{\u\alpha}\,E^{(1)}\Gamma_{11}\chi^{(1)}
\nonumber\\
&&{}
+\frac16E^{(1)\,\u\alpha}\,\Theta\chi^{(1)}
-\frac16(\Gamma_{11}E^{(1)})^{\u\alpha}\,\Theta\Gamma_{11}\chi^{(1)}
-\frac16(\Gamma_a\chi^{(1)})^{\u\alpha}\,E^{(1)}\Gamma^a\Theta
+\frac16(\Gamma_a\Gamma_{11}\chi^{(1)})^{\u\alpha}\,E^{(1)}\Gamma^a\Gamma_{11}\Theta\,.\nonumber
\end{eqnarray}
Using the expressions obtained for the superfields at the linear and quadratic order in $\theta$ we obtain
\begin{eqnarray}
\lefteqn{E^{(3)\,\u\alpha}=
\frac{1}{6}(\mathcal M\mathcal D\Theta)^{\u\alpha}
+\frac{1}{96}e^a\,((M+\tilde M)S\Gamma_a\Theta)^{\u\alpha}
+\frac{1}{96}e^a\,(\Theta\Gamma_a(M+\tilde M)\tilde S)^{\u\alpha}
}
\nonumber\\
&&{}
-\frac{i}{24}e^c\,(\Gamma^{ab}\Theta)^{\u\alpha}\,\Theta\Gamma_cU_{ab}\Theta
+\frac{i}{24}e^c\,(\Gamma^{ab}\Theta)^{\u\alpha}\,\Theta\Gamma_aU_{bc}\Theta
+\frac{i}{24}e^c\,(\Gamma^{ab}\Gamma_{11}\Theta)^{\u\alpha}\,\Theta\Gamma_c\Gamma_{11}U_{ab}\Theta
\nonumber\\
&&{}
+\frac{i}{24}e^c\,(\Gamma^{ab}\Gamma_{11}\Theta)^{\u\alpha}\,\Theta\Gamma_a\Gamma_{11}U_{bc}\Theta
-\frac{i}{48}e^c\,(U_{ab}\Theta)^{\u\alpha}\,\Theta\Gamma_c{}^{ab}\Theta
+\frac{i}{48}e^c\,(\Gamma_{11}U_{ab}\Theta)^{\u\alpha}\,\Theta\Gamma_c{}^{ab}\Gamma_{11}\Theta
\nonumber\\
\end{eqnarray}
where the matrices $\mathcal M$ and $M$ are quadratic in $\Theta$ and were defined in (\ref{eq:M}). Using these expressions in the general expression for the string action (\ref{eq:action}) one obtains the quartic Lagrangian given in (\ref{eq:L4}).

\section{Conclusions}\label{sec:conclusion}
We have constructed the type IIA and type IIB superstring action to quartic order in $\theta$ in a general supergravity background (with vanishing fermionic fields). To obtain this action we had to find the supervielbeins of the background to the same order. This can be done straightforwardly by solving the superspace Bianchi identities order by order in $\theta$ although the expressions quickly become quite involved. We have also argued that knowing the quartic action is especially interesting since for many backgrounds of interest, for example in the AdS/CFT context, one can find a kappa-gauge such that the string action actually truncates at this order. We therefore hope that the action presented here will be useful for semiclassical string computations, especially in the AdS/CFT context. It would be very interesting to study the integrable structure of this action, especially for cases where there is no (complete) supercoset description such as $AdS_2\times S^2\times T^6$. We hope to address this question in the near future.

\vskip .5cm
%%%%%%%%%%%%%%%%%%%%%%%%%%%%%%%%%%%%%%%%%%%%%%%%%%%%%%%%%%%%%%%%%%%%%%%%%%
{\bf Acknowledgements}

I would like to thank Dima Sorokin, Per Sundin and Arkady Tseytlin for useful comments on the draft. This research is supported in part by NSF grant PHY-0906222.

\newpage
\appendix
{\Large{\bf Appendices}}

%%%%%%%%%%%%%%%%%%%%%%%%%%%%%%%%%%%%%%%%%%%%%%%%%%%%%%%%%%%%%%%%%%%%%%%%%%%%%

\section{Spinor and gamma-matrix conventions}\label{app:gamma}
In the type IIA case the Grassmann-odd coordinates are represented as one $32$-component Majorana spinor $\Theta^{\u\alpha}$ ($\u\alpha=1,\ldots,32$). While in the type IIB case they are described as a doublet of $16$-component Majorana-Weyl spinors $\Theta^{\alpha i}$ ($\alpha=1,\ldots,16$ and $i=1,2$).

For the type IIA case the appropriate gamma-matrices are $32\times32$ matrices satisying the Clifford algebra
\begin{equation}
\{\Gamma_a,\Gamma_b\}=2\eta_{ab}\,,
\end{equation}
where the Minkowski metric has mostly plus signature. Together with $\Gamma_{11}=\Gamma_{0\cdots9}$ they form the $D=11$ gamma-matrices $\Gamma_{\hat a}$ ($\hat a=0,\ldots,9,11$). The symmetry properties are as follows
\begin{eqnarray}
\mbox{Symmetric:}&&(C\Gamma_{\hat a})_{\u{\alpha\beta}}\,,\quad (C\Gamma_{\hat a\hat b})_{\u{\alpha\beta}}\,,\quad (C\Gamma_{\hat a\hat b\hat c\hat d\hat e})_{\u{\alpha\beta}}\nonumber\\
\mbox{Anti-symmetric:}&&C_{\u{\alpha\beta}}\,,\quad (C\Gamma_{\hat a\hat b\hat c})_{\u{\alpha\beta}}\,,\quad (C\Gamma_{\hat a\hat b\hat c\hat d})_{\u{\alpha\beta}}\,.
\nonumber
\end{eqnarray}
Here $C$ is the charge conjugation matrix used to raise and lower spinor indices. It satisfies $C^2=-1$. All spinors are defined with an upper index and the charge conjugation matrix will mostly be left implicit, e.g. $\Gamma^a_{\u{\alpha\beta}}=(C\Gamma^a)_{\u{\alpha\beta}}$. It can always be restored by looking at the position of the spinor indices.

The $D=11$ gamma-matrices satisfy the basic Fierz identity
\begin{equation}
\Gamma^{\hat a}_{(\u{\alpha\beta}}(\Gamma_{\hat a\hat b})_{\u{\gamma\delta})}=0\,.
\end{equation}
They also satisfy the duality relations
\begin{eqnarray}
(\Gamma^{a_1\cdots a_{2n}})^{\u\alpha}{}_{\u\beta}&=&\frac{(-1)^n}{(10-2n)!}\varepsilon^{a_1\cdots a_{2n}b_1\cdots b_{10-2n}}(\Gamma_{b_1\cdots b_{10-2n}}\Gamma_{11})^{\u\alpha}{}_{\u\beta}
\nonumber\\
(\Gamma^{a_1\cdots a_{2n+1}})^{\u\alpha}{}_{\u\beta}&=&\frac{(-1)^n}{(9-2n)!}\varepsilon^{a_1\cdots a_{2n+1}b_1\cdots b_{9-2n}}(\Gamma_{b_1\cdots b_{9-2n}}\Gamma_{11})^{\u\alpha}{}_{\u\beta}\,.
\nonumber
\end{eqnarray}

In the type IIB case the appropriate gamma-matrices are instead $16\times16$. They can in fact be taken to be the off-diagonal blocks of the $32\times32$ gamma matrices in the realization
\begin{equation}
(\Gamma^a)^{\u\alpha}{}_{\u\beta}=
\left(\begin{array}{cc}
0 & (\gamma^a)^{\alpha\delta}\\
\gamma^a_{\gamma\beta} & 0
\end{array}\right)\,,
\qquad
(\Gamma_{11})^{\u\alpha}{}_{\u\beta}=
\left(\begin{array}{cc}
\delta^\alpha_\beta & 0\\
0 & -\delta^\delta_\gamma
\end{array}\right)\,,
\qquad
C_{\u{\alpha\beta}}=
\left(\begin{array}{cc}
0 & \delta_\alpha^\delta\\
-\delta^\gamma_\beta &  0
\end{array}\right)\,.
\end{equation}
These have the symmetry properties
\begin{eqnarray}
\mbox{Symmetric:}\qquad\gamma^a_{\alpha\beta}\,,\quad\gamma^{abcde}_{\alpha\beta}\qquad
\mbox{Anti-symmetric:}\qquad\gamma^{abc}_{\alpha\beta}\,.
\nonumber
\end{eqnarray}
In addition they satisfy the basic Fierz identity
\begin{equation}
\gamma^a_{\alpha(\beta}(\gamma_a)_{\gamma\delta)}=0
\end{equation}
and the duality relations
\begin{eqnarray}
(\gamma^{a_1\cdots a_{2n}})^\alpha{}_\beta=\frac{(-1)^n}{(10-2n)!}\varepsilon^{a_1\cdots a_{2n}b_1\cdots b_{10-2n}}(\gamma_{b_1\cdots b_{10-2n}})^\alpha{}_\beta
\nonumber\\
(\gamma^{a_1\cdots a_{2n+1}})_{\alpha\beta}=\frac{(-1)^n}{(9-2n)!}\varepsilon^{a_1\cdots a_{2n+1}b_1\cdots b_{9-2n}}(\gamma_{b_1\cdots b_{9-2n}})_{\alpha\beta}
\nonumber\,.
\end{eqnarray}
Note that $\gamma^{abcde}$ is self-dual.

\section{Type IIB supergravity in superspace}\label{sec:IIB}
Here we briefly review the formulation of type IIB supergravity in superspace \cite{Howe:1983sra} and write the superspace constraints in a useful form for our calculations. The torsion and curvature two-forms are defined as ($A=a,\,\alpha i$)
\begin{eqnarray}
T^A&=&dE^A+E^B\Omega_B{}^A\\
R_A{}^B&=&d\Omega_A{}^B+\Omega_A{}^C\Omega_C{}^B\,,
\end{eqnarray}
in terms of the supervielbeins $E^A$ and spin connection $\Omega^{AB}$, which satisfies
\begin{equation}
\nabla\Theta^{\alpha i}=d\Theta^{\alpha i}+\Omega^\alpha{}_\beta\Theta^{\beta i}
=d\Theta^{\alpha i}-\frac{1}{4}\Omega^{ab}\,(\gamma_{ab}\theta)^{\alpha i}\,,
\qquad
\Omega_\alpha{}^\beta=\Omega^\beta{}_\alpha=-\frac{1}{4}\Omega^{ab}(\gamma_{ab})^\beta{}_\alpha\,.
\end{equation}
The Bianchi identities for the torsion and curvature are
\begin{eqnarray}
dT^A+T^B\Omega_B{}^A&=&E^BR_B{}^A\\
dR_A{}^B+R_A{}^C\Omega_C{}^B-\Omega_A{}^CR_C{}^B&=&0\,.
\end{eqnarray}
In addition we have the NSNS three-form field strength $H$ and RR field strengths $F^{(2n+1)}$ ($n=0,\ldots,4$) with Bianchi identities
\begin{eqnarray}
dH&=&0\\
dF^{(2n+1)}&=&-F^{(2n-1)}H\,.
\end{eqnarray}
Note that we are not using a notation which makes the $SL(2,R)$-invariance of type IIB supergravity manifest \cite{Dall'Agata:1998va,Bergshoeff:2007ma}. Instead we have tried to use simple constraints which look almost identical in the type IIA and type IIB case so as to make it easy to go back and forth between the two.

It is useful to have the Bianchi identities also in components. For the RR-fields, for example, they take the form
\begin{equation}
\nabla_{[A_1}F^{(2n+1)}_{A_2\cdots A_{2n+2}]}+\frac{2n+1}{2}T_{[A_1A_2}{}^BF^{(2n+1)}_{|B|A_3\cdots A_{2n+2}]}
=-\frac{(2n+1)2n}{3!}H_{[A_1A_2A_3}F^{(2n-1)}_{A_4\cdots A_{2n+2}]}\,.
\end{equation}

\subsection{Superspace constraints}
Here we organize the superspace constraints according to the mass-dimension at which they occur. The higher dimension constraints follow from the dimension zero ones (together with certain conventional choices).

\subsubsection*{Dimension 0}
The curvature vanishes at dimension 0 and the non-zero components of the torsion, NSNS three-form $H$ and RR-fields are
\begin{eqnarray}
T_{\alpha i\beta j}{}^c&=&-i\delta_{ij}\gamma^c_{\alpha\beta}\,,\\
H_{\alpha i\beta j c}&=&-i\sigma^3_{ij}(\gamma_c)_{\alpha\beta}\,,\\
F^{(2n+1)}_{\alpha i\beta ja_1\cdots a_{2n-1}}&=&ie^{-\phi}s^n_{ij}(\gamma_{a_1\cdots a_{2n-1}})_{\alpha\beta}\qquad(n=0,\ldots,4)\,,
\end{eqnarray}
where $\phi$ is the dilaton superfield and
\begin{equation}
s^n_{ij}=((\sigma^3)^{n+1}\sigma^1)_{ij}=
\left\{
\begin{array}{cc}
\varepsilon_{ij} & (\mbox{$n$ even})\\
\sigma^1_{ij} & (\mbox{$n$ odd})
\end{array}
\right.\,,
\end{equation}
in terms of the Pauli matrices.

\subsubsection*{Dimension 1/2}
The NSNS three-form $H$ and curvature vanish at dimension 1/2 and the non-zero components of the torsion and RR-fields are
\begin{eqnarray}
T_{\alpha i\beta j}{}^{\gamma k}&=&
\delta_{(\alpha i}^{\gamma k}\chi_{\beta j)}
+(\delta\sigma^3)_{(\alpha i}^{\gamma k}(\sigma^3\chi)_{\beta j)}
-\frac12\delta_{ij}\gamma^a_{\alpha\beta}(\gamma_a\chi)^{\gamma k}
-\frac12\sigma^3_{ij}\gamma^a_{\alpha\beta}(\gamma_a\sigma^3\chi)^{\gamma k}\,,
\\
F^{(2n+1)}_{\alpha ia_1\cdots a_{2n}}&=&-e^{-\phi}s^n_{ij}(\gamma_{a_1\cdots a_{2n}}\chi^j)_\alpha\,,
\end{eqnarray}
where the dilatino superfield is defined as
\begin{eqnarray}
\chi^i_\alpha=\nabla_{\alpha i}\phi\,.
\end{eqnarray}

\subsubsection*{Dimension 1}
The NSNS tree-form $H$ and RR forms $F^{(1)}$ and $F^{(3)}$ are unconstrained at dimension 1. The non-vanishing components of the torsion are
\begin{equation}
T_{a \beta i}{}^{\gamma j}=-\frac{1}{8}H_{abc}\,\sigma^3_{ij}(\gamma^{bc})^\gamma{}_\beta-\frac{1}{8}(S_{ji}\gamma_a)^\gamma{}_\beta\,,
\end{equation}
where the dependence on the RR fields is captured by the anti-symmetric (in $\alpha i\leftrightarrow\beta j$) matrix
\begin{eqnarray}
S^{\alpha\beta}_{ij}
&=&-\frac{1}{2}\sum_{n=0}^4\frac{s^n_{ij}}{(2n+1)!}F'^{(2n+1)}_{b_1\cdots b_{2n+1}}(\gamma^{b_1\cdots b_{2n+1}})^{\alpha\beta}
\nonumber\\
&=&-(F'^{(1)}_a\varepsilon_{ij}\gamma^a+\frac{1}{3!}F'^{(3)}_{abc}\,\sigma^1_{ij}\gamma^{abc}+\frac{1}{2\cdot5!}F'^{(5)}_{abcde}\varepsilon_{ij}\gamma^{abcde})^{\alpha\beta}\,.
\label{eq:SB}
\end{eqnarray}
The modified RR field strengths, denoted with a prime, are defined as
\begin{equation}
F'^{(2n+1)}_{a_1\cdots a_{2n+1}}=e^\phi F^{(2n+1)}_{a_1\cdots a_{2n+1}}+i\chi^1\gamma_{a_1\cdots a_{2n+1}}\chi^2\qquad(n=0,\ldots,4)\,.
\end{equation}
The constraints on the RR-fields at dimension 1 are then the duality relations
\begin{eqnarray}
F'^{(2n-1)}_{a_1\cdots a_{2n-1}}
=\frac{(-1)^{n+1}}{(10-2n+1)!}\varepsilon_{a_1\cdots a_{2n-1}}{}^{a_{2n}\cdots a_{10}}F'^{(10-2n+1)}_{a_{2n}\cdots a_{10}}\qquad(n=3,4,5)\,.
\end{eqnarray}
In particular $F^{(5)}_{abcde}$ is self-dual.

In addition we find at dimension 1 that the spinorial derivative of the dilatino is given by
\begin{eqnarray}
\label{eq:dalphachi}
\nabla_{\alpha i}\chi_\beta^j=
\frac12\chi_{\alpha i}\chi_{\beta j}
+\frac12(\sigma^3\chi)_{\alpha i}(\sigma^3\chi)_{\beta j}
+\frac{i}{2}\nabla_a\phi\,\delta_{ij}\gamma^a_{\alpha\beta}
-\frac{i}{4!}H_{abc}\,\sigma^3_{ij}\gamma^{abc}_{\alpha\beta}-\frac{i}{16}(\gamma^aS_{ij}\gamma_a)_{\alpha\beta}\,.
\end{eqnarray}
The last term is easily computed from the definition of $S$ giving
\begin{equation}
\gamma^aS\gamma_a=4(2F'^{(1)}_a\varepsilon\gamma^a+\frac{1}{3!}F'^{(3)}_{abc}\sigma^1\gamma^{abc})\,.
\end{equation}

Finally, the dimension 1 curvature can be easily found from the torsion Bianchi identity,
\begin{eqnarray}
\label{eq:torsionbianchi}
\nabla_{[A}T_{BC]}{}^D+T_{[AB}{}^ET_{|E|C]}{}^D=R_{[ABC]}{}^D\,,
\end{eqnarray}
which gives at dimension 1
\begin{eqnarray}
R_{\alpha i\beta jc}{}^d
=\nabla_cT_{\alpha i\beta j}{}^d+2T_{c(\alpha i}{}^{\gamma k}T_{\beta j)\gamma k}{}^d
=\frac{i}{2}H_{ac}{}^d\,\sigma^3_{ij}\gamma^a_{\alpha\beta}
-\frac{i}{8}(\gamma_cS_{ij}\gamma^d)_{\alpha\beta}+\frac{i}{8}(\gamma^dS_{ij}\gamma_c)_{\alpha\beta}\,.
\end{eqnarray}

\subsubsection*{Dimension 3/2}
The torsion Bianchi identity (\ref{eq:torsionbianchi}) gives for the dimension 3/2 curvature
\begin{eqnarray}
2R_{\alpha i[bc]d}=-i(\gamma_d\psi^i_{bc})_\alpha\,,
\end{eqnarray}
where we have introduced the gravitino field strength equal to the dimension 3/2 component of the torsion
\begin{equation}
T_{ab}{}^{\alpha i}=\psi_{ab}^{\alpha i}\,.
\end{equation}
This implies that the dimension 3/2 curvature is given by
\begin{equation}
\label{eq:dim-threehalfR}
R_{\alpha ibcd}=\frac{i}{2}(\gamma_b\psi^i_{cd})_\alpha-i(\gamma_{[c}\psi^i_{d]b})_\alpha\,.
\end{equation}

From the Bianchi identity for the NSNS three-form $H$ one finds
\begin{equation}
\label{eq:dalphaH}
\nabla_{\alpha i}H_{abc}=-3T_{[ab}{}^{\beta j}H_{|\alpha i\beta j|c]}=3i(\gamma_{[a}\sigma^3\psi_{bc]})_{\alpha i}\,.
\end{equation}
It will be useful to also have an expression for the spinor derivative of the RR-matrix $S$ defined in (\ref{eq:SB}). This can be derived from the appropriate component of the torsion Bianchi identity (\ref{eq:torsionbianchi}). For $i\neq j$ we have
\begin{eqnarray}
R_{a\beta i\gamma j}{}^{\delta j}
&=&
-\nabla_{\beta i}T_{a\gamma j}{}^{\delta j}
-\nabla_{\gamma j}T_{a\beta i}{}^{\delta j}
+T_{a\beta i}{}^{\eta j}T_{\eta j\gamma j}{}^{\delta j}
\nonumber\\
&=&
-\frac{1}{8}(\gamma_a\nabla_{\gamma j}S_{ij})_\beta{}^\delta
-\frac{1}{8}\sigma^3_{jj}(\gamma^{bc})_\gamma{}^\delta\,\nabla_{\beta i}H_{abc}
+\frac{1}{8}(\gamma_aS_{ij})_\beta{}^\delta\,\chi^j_\gamma
\nonumber\\
&&{}
+\frac{1}{8}\delta_\gamma^\delta\,(\gamma_aS_{ij}\chi^j)_\beta
-\frac{1}{8}(\gamma_aS_{ij}\gamma^b)_{\beta\gamma}(\gamma_b\chi^j)^\delta
\qquad\mbox{($i\neq j$, no sum on $j$)}\,.
\end{eqnarray}
Using the expression for the dimension 3/2 curvature (\ref{eq:dim-threehalfR}) and (\ref{eq:dalphaH}) and multiplying with $\gamma^a$ we get
\begin{eqnarray}
\nabla_{\alpha j}S_{ij}^{\beta\gamma}
&=&
S_{ij}^{\beta\gamma}\,\chi^j_\alpha
+\delta_\alpha^\gamma\,(S_{ij}\chi^j)^\beta
-(S_{ij}\gamma^b)^\beta{}_\alpha\,(\gamma_b\chi^j)^\gamma
+2i(\gamma^{cd})_\alpha{}^\gamma\,(\psi_{cd})^{\beta i}
-2i\delta_{ij}(\gamma^{cd})_\alpha{}^\gamma\,(\psi_{cd})^{\beta j}
\nonumber\\
&&\quad\mbox{(no sum on j)}\,,
\end{eqnarray}
or equivalently,
\begin{eqnarray}
\label{eq:dalphaS}
\nabla_{\alpha i}S^{\beta j\gamma k}
&=&
S^{\beta j\gamma k}\,\chi_{\alpha i}
-\delta_{\alpha i}^{[\beta j}\,(S\chi)^{\gamma k]}
+(\delta\sigma^3)_{\alpha i}^{[\beta j}\,(\sigma^3S\chi)^{\gamma k]}
-(S\gamma^b)^{[\beta j}{}_{\alpha i}\,(\gamma_b\chi)^{\gamma k]}
\nonumber\\
&&{}
+(\sigma^3S\gamma^b)^{[\beta j}{}_{\alpha i}\,(\sigma^3\gamma_b\chi)^{\gamma k]}
-2i(\gamma^{cd})_{\alpha i}{}^{[\beta j}\,(\psi_{cd})^{\gamma k]}
+2i(\sigma^3\gamma^{cd})_{\alpha i}{}^{[\beta j}\,(\sigma^3\psi_{cd})^{\gamma k]}\,.
\nonumber\\
\end{eqnarray}
From the remaining dimension 3/2 torsion and RR-field Bianchi identities one finds that the gamma-trace of the gravitino field strength is given by
\begin{equation}
\gamma^b\psi^i_{ba}=2i\nabla_a\chi^i+\frac{i}{4}H_{abc}\,(\gamma^{bc}\sigma^3\chi)^i
\end{equation}
while the dilatino satisfies the equation of motion
\begin{eqnarray}
\gamma^a\nabla_a\chi^i-2\nabla_a\phi\,\gamma^a\chi^i-\frac{1}{24}H_{abc}\,\sigma^3_{ij}\gamma^{abc}\chi^j+\frac{1}{4}S_{ij}\chi^j=0\,.
\end{eqnarray}

\subsubsection*{Dimension 2}
At dimension 2 we would find the bosonic equations of motion but as we will not need these here we will only derive the expression for the spinorial derivative of the gravitino field strength which we will need. From the torsion Bianchi identity (\ref{eq:torsionbianchi}) we get at dimension 2
\begin{eqnarray}
R_{ab\alpha i}{}^{\beta j}
=
\nabla_{\alpha i}T_{ab}{}^{\beta j}
+2\nabla_{[a}T_{b]\alpha i}{}^{\beta j}
-2T_{\alpha i[a}{}^{\gamma k}T_{b]\gamma k}{}^{\beta j}
+T_{ab}{}^{\gamma k}T_{\gamma k\alpha i}{}^{\beta j}\,.
\end{eqnarray}
This gives the following equation for the spinorial derivative of the gravitino field strength
\begin{eqnarray}
\nabla_{\alpha i}\psi_{ab}{}^{\beta j}
&=&
-\frac{1}{4}\nabla_{[a}H_{b]cd}\,\sigma^3_{ij}(\gamma^{cd})_\alpha{}^\beta
+\frac{1}{4}(\gamma_{[a}\nabla_{b]}S_{ij})_\alpha{}^\beta
+\frac{1}{4}\delta_{ij}(R_{abcd}+\frac{1}{2}H_{ace}H_{bd}{}^e)\,(\gamma^{cd})_\alpha{}^\beta
\nonumber\\
&&{}
-\frac{1}{32}\sigma^3_{ik}H_{cd[a}\,(\gamma^{cd}\gamma_{b]}S_{kj})_\alpha{}^\beta
+\frac{1}{32}\sigma^3_{kj}H_{cd[a}\,(\gamma_{b]}S_{ik}\gamma^{cd})_\alpha{}^\beta
-\frac{1}{32}(\gamma_{[a}S_{ik}\gamma_{b]}S_{kj})_\alpha{}^\beta
\nonumber\\
&&{}
-\frac12\delta_{ij}\delta_\alpha^\beta\,\psi_{ab}\chi
-\frac12\sigma^3_{ij}\delta_\alpha^\beta\,\psi_{ab}\sigma^3\chi
+\frac12\chi_{\alpha i}\,\psi_{ab}^{\beta j}
+\frac12(\sigma^3\chi)_{\alpha i}\,(\sigma^3\psi_{ab})^{\beta j}
\nonumber\\
&&{}
+\frac12(\gamma^c\psi_{ab})_{\alpha i}\,(\gamma_c\chi)^{\beta j}
+\frac12(\gamma^c\sigma^3\psi_{ab})_{\alpha i}\,(\gamma_c\sigma^3\chi)^{\beta j}\,.
\label{eq:dalphapsi}
\end{eqnarray}

\section{Type IIA supergravity in superspace}\label{sec:IIA}
Here we will give a brief description of type IIA supergravity in superspace \cite{Carr:1986tk}. We will use a form of the superspace constraints which is essentially identical to that used for the IIB case. This is useful since it lets us easily go back and forth between the type IIA and type IIB case by just replacing the spinors and gamma matrices appropriately.

The basic definitions of torsion and curvature are the same as in the type IIB case. The spin connection now satisfies
\begin{equation}
\nabla\theta^{\u\alpha}=d\theta^{\u\alpha}+\Omega^{\u\alpha}{}_{\u\beta}\theta^{\u\beta}
=d\theta^{\u\alpha}-\frac{1}{4}\Omega^{ab}(\Gamma_{ab}\theta)^{\u\alpha}
\qquad
\Omega_{\u\alpha}{}^{\u\beta}=\Omega^{\u\beta}{}_{\u\alpha}=-\frac{1}{4}\Omega^{ab}(\Gamma_{ab})^{\u\beta}{}_{\u\alpha}\,.
\end{equation}
The only difference in the field content is that we now have even RR-form field strengths satisfying
\begin{equation}
dF^{(2n)}=-F^{(2n-2)}H\qquad(n=1,\ldots,5)\,,
\end{equation}
or in components
\begin{equation}
\nabla_{[A_1}F^{(2n)}_{A_2\cdots A_{2n+1}]}+\frac{2n}{2}T_{[A_1A_2}{}^BF^{(2n)}_{|B|A_3\cdots A_{2n+1}]}
=-\frac{2n(2n-1)}{3!}H_{[A_1A_2A_3}F^{(2n-2)}_{A_4\cdots A_{2n+1}]}\,.
\end{equation}
We take the scalar field strength $F^{(0)}$ to vanish, i.e. we will not consider the Romans massive type IIA supergravity.

Since the calculations are essentially identical to those of the IIB case we will just list the superspace constraints for the IIA case.

\subsection{Superspace constraints}

\subsubsection*{Dimension 0}
The non-zero components at dimension 0 are
\begin{eqnarray}
T_{\u{\alpha\beta}}{}^c&=&-i\Gamma^c_{\u{\alpha\beta}}\,,\\
H_{\u{\alpha\beta}c}&=&-i(\Gamma_c\Gamma_{11})_{\u{\alpha\beta}}\,,
\label{eq:dimzeroH}\\
F^{(2n)}_{\u{\alpha\beta}a_1\cdots a_{2n-2}}&=&ie^{-\phi}(\Gamma_{a_1\cdots a_{2n-2}}(-\Gamma_{11})^n)_{\u{\alpha\beta}}\qquad(n=1,\ldots,5)\,.
\end{eqnarray}

\subsubsection*{Dimension 1/2}
The non-zero components at dimension 1/2 are
\begin{eqnarray}
T_{\u{\alpha\beta}}{}^{\u\gamma}&=&
\delta_{(\u\alpha}^{\u\gamma}\chi_{\u\beta)}
-(\Gamma_{11})^{\u\gamma}{}_{(\u\alpha}(\Gamma_{11}\chi)_{\u\beta)}
-\frac12\Gamma^a_{\u{\alpha\beta}}(\Gamma_a\chi)^{\u\gamma}
+\frac12(\Gamma^a\Gamma_{11})_{\u{\alpha\beta}}(\Gamma_a\Gamma_{11}\chi)^{\u\gamma}\,,
\label{eq:TalphaAhalf}
\\
F^{(2n)}_{\u\alpha a_1\cdots a_{2n-1}}&=&e^{-\phi}(\Gamma_{a_1\cdots a_{2n-1}}\Gamma_{11}^n\chi)_{\u\alpha}\,,
\end{eqnarray}
where the dilatino superfield is defined as $\chi^{\u\alpha}=-C^{\u{\alpha\beta}}\nabla_{\u\beta}\phi$.

\subsubsection*{Dimension 1}
The non-zero components at dimension 1 are
\begin{eqnarray}
T_{a\u\beta}{}^{\u\gamma}&=&-\frac{1}{8}H_{abc}\,(\Gamma^{bc}\Gamma_{11})^{\u\gamma}{}_{\u\beta}-\frac{1}{8}(S\Gamma_a)^{\u\gamma}{}_{\u\beta}\,,
\label{eq:TalphaAone}
\\
R_{\u{\alpha\beta}cd}&=&\frac{i}{2}H_{acd}\,(\Gamma^a\Gamma_{11})_{\u{\alpha\beta}}-\frac{i}{4}(\Gamma_{[c}S\Gamma_{d]})_{\u{\alpha\beta}}\,,
\label{eq:RAone}
\end{eqnarray}
where
\begin{equation}
\label{eq:SA}
S=F'^{(0)}+\frac12F'^{(2)}_{ab}\Gamma^{ab}\Gamma_{11}+\frac{1}{4!}F'^{(4)}_{abcd}\Gamma^{abcd}\,.
\end{equation}
The modified RR field strengths, denoted with a prime, are defined as
\begin{equation}
F'^{(2n)}_{a_1\cdots a_{2n}}=e^\phi F^{(2n)}_{a_1\cdots a_{2n}}+\frac{i}{2}\chi\Gamma_{a_1\cdots a_{2n}}\Gamma_{11}^n\chi\qquad(n=0,...,5)\,.
\end{equation}
Note that $F^{(0)}$, the Romans mass parameter, is taken to vanish but $F'^{(0)}$ is still non-zero. The constraints on the RR-fields at dimension 1 are then the duality relations
\begin{eqnarray}
F'^{(2n)}_{a_1\cdots a_{2n}}
=\frac{(-1)^{n+1}}{(10-2n)!}\varepsilon_{a_1\cdots a_{2n}}{}^{a_{2n+1}\cdots a_{10}}F'^{(10-2n)}_{a_{2n+1}\cdots a_{10}}\qquad(n=3,4,5)\,.
\end{eqnarray}

The spinorial derivative of the dilatino is given by
\begin{eqnarray}
\nabla_{\u\alpha}\chi^{\u\beta}
=
\frac12\chi_{\u\alpha}\chi^{\u\beta}
+\frac12(\Gamma_{11}\chi)_{\u\alpha}(\Gamma_{11}\chi)^{\u\beta}
+\frac{i}{2}\nabla_a\phi\,(\Gamma^a)^{\u\beta}{}_{\u\alpha}
+\frac{i}{4!}H_{abc}\,(\Gamma^{abc}\Gamma_{11})^{\u\beta}{}_{\u\alpha}
+\frac{i}{16}(\Gamma^aS\Gamma_a)^{\u\beta}{}_{\u\alpha}\,.
\nonumber\\
\label{eq:dalphachiA}
\end{eqnarray}

\subsubsection*{Dimension 3/2}
At dimension 3/2 we have
\begin{eqnarray}
\label{eq:dim-threehalfRA}
R_{\u\alpha bcd}&=&\frac{i}{2}(\Gamma_b\psi_{cd})_{\u\alpha}-i(\Gamma_{[c}\psi_{d]b})_{\u\alpha}\,,
\\
\label{eq:dalphaHA}
\nabla_{\u\alpha}H_{abc}&=&3i(\Gamma_{[a}\Gamma_{11}\psi_{bc]})_{\u\alpha}\,,
\end{eqnarray}
where the gravitino field strength is defined as $\psi_{ab}^{\u\alpha}=T_{ab}{}^{\u\alpha}$. We also have
\begin{eqnarray}
\lefteqn{\nabla_{\u\alpha}S^{\u{\beta\gamma}}=
S^{\u{\beta\gamma}}\,\chi_{\u\alpha}
+\delta^{[\u\beta}_{\u\alpha}\,(S\chi)^{\u\gamma]}
-(\Gamma_{11})^{[\u\beta}{}_{\u\alpha}\,(\Gamma_{11}S\chi)^{\u\gamma]}
+(S\Gamma^b)^{[\u\beta}{}_{\u\alpha}\,(\Gamma_b\chi)^{\u\gamma]}
}
\nonumber\\
&&{}
-(\Gamma_{11}S\Gamma^b)^{[\u\beta}{}_{\u\alpha}\,(\Gamma_{11}\Gamma_b\chi)^{\u\gamma]}
-2i(\Gamma^{cd})^{[\u\beta}{}_{\u\alpha}\,(\psi_{cd})^{\u\gamma]}
+2i(\Gamma^{cd}\Gamma_{11})^{[\u\beta}{}_{\u\alpha}\,(\Gamma_{11}\psi_{cd})^{\u\gamma]}\,.
\label{eq:dalphaSA}
\end{eqnarray}
As well as
\begin{equation}
\Gamma^b\psi_{ba}=2i\nabla_a\chi-\frac{i}{4}H_{abc}\,\Gamma^{bc}\Gamma_{11}\chi
\end{equation}
and the dilatino equation of motion
\begin{eqnarray}
\Gamma^a\nabla_a\chi-2\nabla_a\phi\,\Gamma^a\chi+\frac{1}{24}H_{abc}\,\Gamma^{abc}\Gamma_{11}\chi+\frac{1}{4}S\chi=0\,.
\end{eqnarray}

\subsubsection*{Dimension 2}
The spinorial derivative of the gravitino field strength is given by
\begin{eqnarray}
\nabla_{\u\alpha}\psi_{ab}{}^{\u\beta}
&=&
\frac{1}{4}\nabla_{[a}H_{b]cd}\,(\Gamma^{cd}\Gamma_{11})^{\u\beta}{}_{\u\alpha}
+\frac{1}{4}(\nabla_{[a}S\Gamma_{b]})^{\u\beta}{}_{\u\alpha}
-\frac{1}{4}(R_{abcd}+\frac12H_{ace}H_{bd}{}^e)(\Gamma^{cd})^{\u\beta}{}_{\u\alpha}
\nonumber\\
&&{}
-\frac{1}{32}H_{cd[a}\,(S\Gamma_{b]}\Gamma^{cd}\Gamma_{11})^{\u\beta}{}_{\u\alpha}
+\frac{1}{32}H_{cd[a}\,(\Gamma^{cd}\Gamma_{11}S\Gamma_{b]})^{\u\beta}{}_{\u\alpha}
+\frac{1}{32}(S\Gamma_{[a}S\Gamma_{b]})^{\u\beta}{}_{\u\alpha}
\nonumber\\
&&{}
-\frac12\delta_{\u\alpha}^{\u\beta}\,\psi_{ab}\chi
+\frac12(\Gamma_{11})^{\u\beta}{}_{\u\alpha}\,\psi_{ab}\Gamma_{11}\chi
+\frac12\chi_{\u\alpha}\,\psi_{ab}^{\u\beta}
-\frac12(\Gamma_{11}\chi)_{\u\alpha}\,(\Gamma_{11}\psi_{ab})^{\u\beta}
\nonumber\\
&&{}
+\frac12(\Gamma^c\psi_{ab})_{\u\alpha}(\Gamma_c\chi)^{\u\beta}
-\frac12(\Gamma^c\Gamma_{11}\psi_{ab})_{\u\alpha}(\Gamma_c\Gamma_{11}\chi)^{\u\beta}\,.
\label{eq:dalphapsiA}
\end{eqnarray}


\begin{thebibliography}{20}

%\cite{Maldacena:1997re}
\bibitem{Maldacena:1997re}
  J.~M.~Maldacena,
  ``The Large N limit of superconformal field theories and supergravity,''
  Adv.\ Theor.\ Math.\ Phys.\  {\bf 2} (1998) 231
  [hep-th/9711200].
  %%CITATION = HEP-TH/9711200;%%
  %8927 citations counted in INSPIRE as of 19 Apr 2013
  
%\cite{Bena:2003wd}
\bibitem{Bena:2003wd}
  I.~Bena, J.~Polchinski and R.~Roiban,
  ``Hidden symmetries of the AdS(5) x S**5 superstring,''
  Phys.\ Rev.\ D {\bf 69} (2004) 046002
  [hep-th/0305116].
  %%CITATION = HEP-TH/0305116;%%
  %583 citations counted in INSPIRE as of 19 Apr 2013

%\cite{Arutyunov:2008if}
\bibitem{Arutyunov:2008if}
  G.~Arutyunov and S.~Frolov,
  ``Superstrings on AdS(4) x CP**3 as a Coset Sigma-model,''
  JHEP {\bf 0809} (2008) 129
  [arXiv:0806.4940 [hep-th]].
  %%CITATION = ARXIV:0806.4940;%%
  %168 citations counted in INSPIRE as of 19 Apr 2013

%\cite{Stefanski:2008ik}
\bibitem{Stefanski:2008ik}
  B.~Stefanski, jr,
  ``Green-Schwarz action for Type IIA strings on AdS(4) x CP**3,''
  Nucl.\ Phys.\ B {\bf 808} (2009) 80
  [arXiv:0806.4948 [hep-th]].
  %%CITATION = ARXIV:0806.4948;%%
  %159 citations counted in INSPIRE as of 19 Apr 2013

%\cite{Gomis:2008jt}
\bibitem{Gomis:2008jt}
  J.~Gomis, D.~Sorokin and L.~Wulff,
  ``The Complete AdS(4) x CP**3 superspace for the type IIA superstring and D-branes,''
  JHEP {\bf 0903} (2009) 015
  [arXiv:0811.1566 [hep-th]].
  %%CITATION = ARXIV:0811.1566;%%
  %91 citations counted in INSPIRE as of 19 Apr 2013

%\cite{Cagnazzo:2012se}
\bibitem{Cagnazzo:2012se}
  A.~Cagnazzo and K.~Zarembo,
  ``B-field in AdS(3)/CFT(2) Correspondence and Integrability,''
  JHEP {\bf 1211} (2012) 133
   [Erratum-ibid.\  {\bf 1304} (2013) 003]
  [arXiv:1209.4049 [hep-th]].
  %%CITATION = ARXIV:1209.4049;%%
  %11 citations counted in INSPIRE as of 19 Apr 2013
  
%\cite{Babichenko:2009dk}
\bibitem{Babichenko:2009dk}
  A.~Babichenko, B.~Stefanski, Jr. and K.~Zarembo,
  ``Integrability and the AdS(3)/CFT(2) correspondence,''
  JHEP {\bf 1003} (2010) 058
  [arXiv:0912.1723 [hep-th]].
  %%CITATION = ARXIV:0912.1723;%%
  %59 citations counted in INSPIRE as of 19 Apr 2013
  
%\cite{Sorokin:2011rr}
\bibitem{Sorokin:2011rr}
  D.~Sorokin, A.~Tseytlin, L.~Wulff and K.~Zarembo,
  ``Superstrings in AdS(2)xS(2)xT(6),''
  J.\ Phys.\ A {\bf 44} (2011) 275401
  [arXiv:1104.1793 [hep-th]].
  %%CITATION = ARXIV:1104.1793;%%
  %23 citations counted in INSPIRE as of 19 Apr 2013


%\cite{Sorokin:2010wn}
\bibitem{Sorokin:2010wn}
  D.~Sorokin and L.~Wulff,
  ``Evidence for the classical integrability of the complete $AdS_4 x CP^3$ superstring,''
  JHEP {\bf 1011} (2010) 143
  [arXiv:1009.3498 [hep-th]].
  %%CITATION = ARXIV:1009.3498;%%
  %26 citations counted in INSPIRE as of 19 Apr 2013

%\cite{Sundin:2012gc}
\bibitem{Sundin:2012gc}
  P.~Sundin and L.~Wulff,
  ``Classical integrability and quantum aspects of the AdS(3) x S(3) x S(3) x S(1) superstring,''
  JHEP {\bf 1210} (2012) 109
  [arXiv:1207.5531 [hep-th]].
  %%CITATION = ARXIV:1207.5531;%%
  %14 citations counted in INSPIRE as of 19 Apr 2013

%\cite{Metsaev:1998it}
\bibitem{Metsaev:1998it}
  R.~R.~Metsaev and A.~A.~Tseytlin,
  ``Type IIB superstring action in AdS(5) x S**5 background,''
  Nucl.\ Phys.\ B {\bf 533} (1998) 109
  [hep-th/9805028].
  %%CITATION = HEP-TH/9805028;%%
  %494 citations counted in INSPIRE as of 12 Apr 2013

%\cite{Russo:1998xv}
\bibitem{Russo:1998xv}
  J.~G.~Russo and A.~A.~Tseytlin,
  ``Green-Schwarz superstring action in a curved magnetic Ramond-Ramond background,''
  JHEP {\bf 9804} (1998) 014
  [hep-th/9804076].
  %%CITATION = HEP-TH/9804076;%%
  %30 citations counted in INSPIRE as of 12 Apr 2013

%\cite{Cvetic:1999zs}
\bibitem{Cvetic:1999zs}
  M.~Cvetic, H.~Lu, C.~N.~Pope and K.~S.~Stelle,
  ``T duality in the Green-Schwarz formalism, and the massless / massive IIA duality map,''
  Nucl.\ Phys.\ B {\bf 573} (2000) 149
  [hep-th/9907202].
  %%CITATION = HEP-TH/9907202;%%
  %96 citations counted in INSPIRE as of 12 Apr 2013

%\cite{Tsimpis:2004gq}
\bibitem{Tsimpis:2004gq}
  D.~Tsimpis,
  ``Curved 11D supergeometry,''
  JHEP {\bf 0411} (2004) 087
  [hep-th/0407244].
  %%CITATION = HEP-TH/0407244;%%
  %17 citations counted in INSPIRE as of 19 Apr 2013
  
%\cite{Sahakian:2004gy}
\bibitem{Sahakian:2004gy}
  V.~Sahakian,
  ``Closed strings in Ramond-Ramond backgrounds,''
  JHEP {\bf 0404} (2004) 026
  [hep-th/0402037].
  %%CITATION = HEP-TH/0402037;%%
  %6 citations counted in INSPIRE as of 12 Apr 2013
  
%\cite{Grisaru:1985fv}
\bibitem{Grisaru:1985fv}
  M.~T.~Grisaru, P.~S.~Howe, L.~Mezincescu, B.~Nilsson and P.~K.~Townsend,
  ``N=2 Superstrings in a Supergravity Background,''
  Phys.\ Lett.\ B {\bf 162} (1985) 116.
  %%CITATION = PHLTA,B162,116;%%
  %130 citations counted in INSPIRE as of 12 Apr 2013
  
%\cite{Tseytlin:1996hs}
\bibitem{Tseytlin:1996hs}
  A.~A.~Tseytlin,
  ``On dilaton dependence of type II superstring action,''
  Class.\ Quant.\ Grav.\  {\bf 13} (1996) L81
  [hep-th/9601109].
  %%CITATION = HEP-TH/9601109;%%
  %18 citations counted in INSPIRE as of 12 Apr 2013

%\cite{Metsaev:2000yf}
\bibitem{Metsaev:2000yf}
  R.~R.~Metsaev and A.~A.~Tseytlin,
  ``Superstring action in AdS(5) x S**5. Kappa symmetry light cone gauge,''
  Phys.\ Rev.\ D {\bf 63} (2001) 046002
  [hep-th/0007036].
  %%CITATION = HEP-TH/0007036;%%
  %99 citations counted in INSPIRE as of 15 May 2013

%\cite{Metsaev:2000yu}
\bibitem{Metsaev:2000yu}
  R.~R.~Metsaev, C.~B.~Thorn and A.~A.~Tseytlin,
  ``Light cone superstring in AdS space-time,''
  Nucl.\ Phys.\ B {\bf 596} (2001) 151
  [hep-th/0009171].
  %%CITATION = HEP-TH/0009171;%%
  %113 citations counted in INSPIRE as of 15 May 2013
  
%\cite{Atick:1986jr}
\bibitem{Atick:1986jr}
  J.~J.~Atick and A.~Dhar,
  ``Normal Coordinates, Theta Expansion And Strings On Curved Superspace,''
  Nucl.\ Phys.\ B {\bf 284} (1987) 131.
  %%CITATION = NUPHA,B284,131;%%
  %30 citations counted in INSPIRE as of 12 Apr 2013

%\cite{Kallosh:1998zx}
\bibitem{Kallosh:1998zx}
  R.~Kallosh, J.~Rahmfeld and A.~Rajaraman,
  ``Near horizon superspace,''
  JHEP {\bf 9809} (1998) 002
  [hep-th/9805217].
  %%CITATION = HEP-TH/9805217;%%
  %128 citations counted in INSPIRE as of 15 May 2013
  
%\cite{deWit:1998yu}
\bibitem{deWit:1998yu}
  B.~de Wit, K.~Peeters, J.~Plefka and A.~Sevrin,
  ``The M theory two-brane in AdS(4) x S**7 and AdS(7) x S**4,''
  Phys.\ Lett.\ B {\bf 443} (1998) 153
  [hep-th/9808052].
  %%CITATION = HEP-TH/9808052;%%
  %68 citations counted in INSPIRE as of 18 Apr 2013

%\cite{Howe:1983sra}
\bibitem{Howe:1983sra}
  P.~S.~Howe and P.~C.~West,
  ``The Complete N=2, D=10 Supergravity,''
  Nucl.\ Phys.\ B {\bf 238} (1984) 181.
  %%CITATION = NUPHA,B238,181;%%
  %389 citations counted in INSPIRE as of 12 Apr 2013

%\cite{Dall'Agata:1998va}
\bibitem{Dall'Agata:1998va}
  G.~Dall'Agata, K.~Lechner and M.~Tonin,
  ``D = 10, N = IIB supergravity: Lorentz invariant actions and duality,''
  JHEP {\bf 9807} (1998) 017
  [hep-th/9806140].
  %%CITATION = HEP-TH/9806140;%%
  %84 citations counted in INSPIRE as of 20 Apr 2013
  
%\cite{Bergshoeff:2007ma}
\bibitem{Bergshoeff:2007ma}
  E.~Bergshoeff, P.~S.~Howe, S.~Kerstan and L.~Wulff,
  ``Kappa-symmetric SL(2,R) covariant D-brane actions,''
  JHEP {\bf 0710} (2007) 050
  [arXiv:0708.2722 [hep-th]].
  %%CITATION = ARXIV:0708.2722;%%
  %9 citations counted in INSPIRE as of 19 Apr 2013
  
  
%\cite{Carr:1986tk}
\bibitem{Carr:1986tk}
  J.~L.~Carr, S.~J.~Gates, Jr. and R.~N.~Oerter,
  ``D = 10, N=2a Supergravity in Superspace,''
  Phys.\ Lett.\ B {\bf 189} (1987) 68.
  %%CITATION = PHLTA,B189,68;%%
  %32 citations counted in INSPIRE as of 12 Apr 2013
    
\end{thebibliography}
\end{document}